\documentclass{aa}  

\usepackage{hyperref}
\usepackage{graphicx}
\usepackage[version=3]{mhchem} % Formula subscripts using \ce{}
\usepackage{xcolor}
\usepackage{cleveref}
\usepackage{booktabs}
\usepackage{txfonts}
\usepackage{ulem}

\begin{document}

   \title{Cracking the Puzzle of \ce{CO2} Formation on Interstellar Ices}

   \subtitle{Quantum Chemical and Kinetic Study of the \ce{CO + OH -> CO2 + H} Reaction.}

   \author{G. Molpeceres
          \inst{1}
          \and
          J. Enrique-Romero
          \inst{2}
          \and
          Y. Aikawa
          \inst{1}
          }

   \institute{Department of Astronomy, Graduate School of Science, The University of Tokyo, Tokyo 113 0033, Japan\\
              \email{molpeceres@astron.s.u-tokyo.ac.jp}
         \and
             Leiden Institute of Chemistry, Gorlaeus Laboratories, Leiden University, PO Box 9502, 2300 RA Leiden, The Netherlands \\
             \email{j.enrique.romero@lic.leidenuniv.nl}
             }

   \date{Received \today; accepted \today}

% \abstract{}{}{}{}{} 
% 5 {} token are mandatory
 
  \abstract
  % context heading (optional)
  % {} leave it empty if necessary  
   {\ce{CO2} is one of the dominant components of the interstellar ice. Recent observations show \ce{CO2} exists more abundantly in polar (\ce{H2O}-dominated) ice than in apolar (\ce{H2O}-poor) ice. \ce{CO2} ice formation is primarily attributed to the reaction between CO and OH, which has a barrier.}
  % aims heading (mandatory)
   {We investigate the title reaction in \ce{H2O} ice and CO ice to quantify the efficiency of the reaction in polar ice and apolar ice.}
  % methods heading (mandatory)
   {Highly accurate quantum chemical calculations were employed to analyze the stationary points of the potential energy surfaces of the title reaction in the gas phase on a \ce{H2O} and CO clusters. Microcanonical transition state theory was used as a diagnostic tool for the efficiency of the reaction under ISM conditions. We simulate the kinetics of ice chemistry, considering different scenarios involving non-thermal processes and energy dissipation.}
  % results heading (mandatory)
   {The CO + OH reaction proceeds through the remarkably stable intermediate HOCO radical. On the \ce{H2O} cluster, the formation of this intermediate is efficient, but the subsequent reaction leading to \ce{CO2} formation is not. Conversely, HOCO formation on the CO cluster is inefficient without external energy input. Thus, \ce{CO2} ice cannot be formed by the title reaction alone either on \ce{H2O} cluster or CO cluster.}
  % conclusions heading (optional), leave it empty if necessary 
   {In the polar ice, \ce{CO2} ice formation is possible via \ce{CO + OH -> HOCO}, followed by \ce{HOCO + H ->CO2 + H2}, as demonstrated by abundant experimental literature. In apolar ice, \ce{CO2} formation is less efficient because HOCO formation requires external energy. Our finding is consistent with the JWST observations. Further experimental work is encouraged using low-temperature OH radicals.}


   \maketitle
%
%-------------------------------------------------------------------

\section{Introduction} \label{sec:introduction}

In the cold molecular clouds of the interstellar medium (ISM), a significant fraction of the molecules are contained in the solid phase in the form of ice. While most of the molecules present in the ISM have been detected in the gas phase using radio telescopes through their rotational transitions, the direct observation of ices requires studying their vibrational transitions, which are commonly affected by telluric contamination. In this context, space telescopes, such as Spitzer or, more recently, JWST, are essential. Ice observations \citet{Oberg2011, Boogert2015, McClure2023} reveal the presence of several components such as \ce{H2O}, \ce{CO}, \ce{CH3OH}, and the object of this study, \ce{CO2}. The abundance of these species, as well as their speciation in the ice or their presence in specific regions of the ISM, can only be explained by considering their formation routes and the chemical conditions necessary for their appearance. 

The different components of interstellar ice may be formed in situ on the surface of refractory material. Such is the case of \ce{H2O}, which is formed from the hydrogenation of atomic oxygen \citep{ioppolo_laboratory_2008, Dulieu2010,Lamberts2014, lamberts_water_2013, Meisner2017, Molpeceres2019}, or the case of \ce{CH3OH}, which is formed from the hydrogenation of CO \citep{Watanabe2002, Fuchs2009, Rimola2014}. Other significant components are primarily synthesized in the gas and accrete under extremely cold and dense conditions on the grain, like CO. Interstellar carbon dioxide, \ce{CO2}, is thought to form via reactions on the surface (see, e.g., \citet{Garrod2011, Pauly2018}). The postulated reactions contributing to the \ce{CO2} formation are:

\begin{align}
    \ce{CO + OH &-> CO2 + H} \label{reac:1}\\
    \ce{HCO + O &-> CO2 + H} \label{reac:2} \\
    \ce{CO + O &-> CO2} \label{reac:3}
\end{align}

From this ternary of reactions, Reaction \ref{reac:3} has a barrier energy when atomic oxygen is in its ground state, ($^{3}$P)O \citep{minissale_co_2013}. Reaction \ref{reac:2} is barrierless, and Reaction \ref{reac:1}, the reaction whose study we tackle in this paper, is assumed to have a minimal activation energy ($\sim$ 100 K, \citet{Garrod2011}. 

The assumption of tiny activation energy for the \ce{CO + OH -> CO2 + H} reaction is supported by a plethora of experiments  dealing with surface chemical experiments \citep{ioppolo_surface_2011, noble_co_2011, oba_experimental_2010, Oba2010Carbonic, Qasim2019, 2021GutierrezQuinanilla, terwisscha_van_scheltinga_formation_2022}. Each of these experiments vary in different factors, including the formation route of the OH radical, either by hydrogenation of \ce{O2}, \citep{ioppolo_surface_2011, noble_co_2011, Qasim2019}, dissociation of \ce{H2O} molecules before deposition on the ice \citep{oba_experimental_2010, Oba2010Carbonic}, or direct photodissociation of \ce{H2O} ice molecules \citep{terwisscha_van_scheltinga_formation_2022}. Other variations between experiments include the substrate under consideration, either amorphous silicates \citep{noble_co_2011}, \ce{CO} \citep{Qasim2019}, matrix isolation \cite{2021GutierrezQuinanilla} or \ce{H2O} \citep{ioppolo_surface_2011, oba_experimental_2010, Oba2010Carbonic, terwisscha_van_scheltinga_formation_2022}. On the modelling side, \citet{Garrod2011} build on the experimental knowledge and coarse-grained it in a combination of a direct formation route \ce{CO + OH -> CO2 + H} operating at $T$$\geq$12 K, coinciding with the onset of CO diffusion on \ce{H2O}, and an indirect three-body route on CO ices that relies in the formation of a kinetically excited OH radical \ce{O + H -> OH$^{*}$} that subsequently partakes in the \ce{CO + OH$^{*}$} reaction. The latter route on CO ices allows to explain the \ce{CO2} bands in a non-polar media observed in infrared observations of ices \citep{Oberg2011, Boogert2015, McClure2023}. In summary, there is ample evidence for Reaction \ref{reac:1}, to be efficient on dust grains. However, the same reaction in the gas phase is relatively slow, with rate constants as low as $\sim$ 2x10$^{-13}$ molecules cm$^{-3}$ s$^{-1}$ at 300 K \citep{Frost1991}. The title reaction in the gas phase has also been a source of extensive theoretical attention. It has been simulated using both semi-classical and quantum dynamics on highly accurate potential energy surfaces (PES) \citep{Ma2012, caracciolo_combined_2018, li_quantum_2014}. It was also studied in the presence of other \ce{CO2} molecules \cite{masunov_catalytic_2018}. The theoretical works find rate constants even lower than the values reported in \citet{Frost1991}. 

The different reactivity on surfaces and the gas phase is puzzling and counterintuitive. In both phases, the reaction is acknowledged to proceed through the highly stable HOCO radical. The evolution from this radical is the primary source of uncertainty because of the high activation energies to form the bimolecular \ce{CO2 + H} products. In the gas, where a third body to stabilize HOCO is unavailable, the reaction is more likely to occur owing to the energy redistribution into the few vibrational degrees of freedom, ultimately leading to an irreversible reaction. On the surface, the ice molecules dissipate a significant fraction of this energy, ideally leading to the thermalization of HOCO, hence slowing or impeding the formation of \ce{CO2}. This was proved by \citet{arasa_molecular_2013}, initiating the conundrum we tackle in this work and that has also been debated from different prisms \citep{bredehoft_co2_2020,Upadhyay2021, tachikawa_reactions_2021}. If the reaction is slow in the gas, it should not proceed on the ice, where little energy is left for the reaction after dissipation into the ice. Hence, how is the mismatch between gas and solid phase experiments possible? In this article, we aim to shed light on this particular issue. The two main possibilities to explain the disagreement include, in the first place, the operation of external energy input, either chemical from the \ce{O2 + H} or \ce{O + H} reactions required to form the OH radical, or the excess energy used to photodissociate \ce{H2O}. Secondly, free H atoms from the experiment may promote H abstraction reactions, \ce{HOCO + H -> CO2 + H2}. While these two possibilities are often assumed when interpreting the experimental results, it is fundamental to distinguish which is dominant, if any, to establish under which conditions the laboratory measurements apply to the ISM. Determining the factors contributing to the reaction yield in the experiments is complicated because the detection techniques are suited for identifying only the final products. Quantum chemical calculations are instrumental and provide an atomistic perspective of the different elementary processes relevant to the reaction. 

In this work, we simulate the title reaction on two different model ices, \ce{H2O} and \ce{CO}, and perform kinetic simulations using a microcanonical formalism to determine the importance of non-thermal effects in the reaction, including dissipation over different numbers of molecules, and complete the picture left by the different experimental studies.
The paper is structured as follows. In \Cref{sec:methods}, we describe the employed computational methodology. In \Cref{sec:results} we present the structural models for the ices (\Cref{sec:res:ices}), the PES for the reactions in each of the surfaces (\Cref{sec:pes} and \Cref{sec:pesonCO}) and the associated kinetic analysis (\Cref{sec:rrkm}). \Cref{sec:discussion} is dedicated to interpreting our results from an astrophysical point of view, contextualising the preceding experiments. We finally summarize our main findings in \Cref{sec:conclusions}.

%--------------------------------------------------------------------
\section{Methodology} \label{sec:methods}

\subsection{Quantum chemical calculations} \label{sec:quantum}

The stationary points in the PES were characterized using density functional theory (DFT) calculations on model clusters mimicking \ce{H2O} and \ce{CO} ices. Because this work aims to determine the impact of energy redistribution in the formation of \ce{CO2} on ice, we need to use sufficiently large structural models to allow for (ergodic) energy equipartition. In a preceding calculation, \cite{Rimola2018} used a cluster containing 33 \ce{H2O} water molecules and discussed the suitability of a model of this size, indicating that energy dissipation should be well described with a model of this size. This was later confirmed with dedicated studies using ab-initio molecular dynamics simulations \citep{Pantaleone2021, Pantaleone2020, Ferrero2023, molpeceres_reaction_2023}. Therefore, in this study, we use the same 33 \ce{H2O} cluster to simulate the \ce{H2O} ice \citep{Rimola2018}, and we constructed a 33 CO cluster to simulate the CO ice. To construct such a cluster, we used {\sc Packmol } \citep{Packmol2009} in a 8 \AA{} radius sphere, ensuring that every molecule is at a minimum initial distance of 3 \AA{} from each other. This initial cluster is later refined at the level of the theory described below. 

The geometries of the initial clusters were optimized at the MN15-D3BJ/6-31+G(d,p) level of theory \citep{mn15, Grimme2010, Grimme2011, hehre1972a, hariharan1973a, ditchfield1971a, clark1983a}, with parameters for the D3BJ dispersion correction taken from \citet{Goerigk2017}. The DFT and optimizations utilize the {\sc Gaussian16} (rev.C.01) suite of programs \citep{g16}. We later place the CO and OH admolecules on the clusters sequentially, first occupying a binding site for the CO molecule and later for OH. Once the two admolecules are located on the clusters, we followed the gas-phase reaction mechanism presented in \citet{Ma2012} for both clusters, except for an alternative exit path on CO ice (\Cref{sec:pesonCO}). Additional differences between the gas-phase and surface-like profiles are highlighted in \Cref{sec:pes}. After locating every stationary point, we confirmed them as either true minima or first-order saddle points, i.e., transition states (TS), in the PES by computing the molecular Hessian of the system. The electronic energies of the stationary points on the PES were further refined using the domain-based local pair-natural orbital coupled cluster singles and doubles with a perturbative treatment of triple excitations, DLPNO-CCSD(T) \citep{riplinger_sparse_2016, guo_communication_2018} using a two-point complete basis set extrapolation (CBS) to the basis-set limit using the cc-pVDZ and cc-pVTZ basis sets \citep{Woon1993, Helgaker1997,Neese2011, Zhong2008, Neese2020}. The internal options for the PNO localization scheme were set to normal, and resolution of the identity (RI) techniques were used to evaluate exchange and Coulomb integrals (RIJK) using a cc-PVTZ/JK auxiliary basis set. We apply the frozen-core approximation in the correlated calculations. The ORCA (v.5.0.4) code was used for the DLPNO-CCSD(T)/CBS calculations \citep{Neese2012,Neese2020, Neese2022}. 

In addition to cluster calculations, we also carried out gas-phase calculations at the same level of theory for comparative purposes, which are indicated throughout the paper in square brackets. Finally, we assessed the quality of our theoretical method of choice, comparing our gas phase results with the ones of \citet{Ma2012}, finding excellent agreement for all the relevant parts of the PES. These results are presented in the \Cref{sec:appendix1}. It is worth noting here that our theoretical method does not predict the correct energetics for the high energy intermediate \ce{HCO2}. This intermediate is not relevant to the kinetics of the system because its formation requires surmounting an emerged barrier of $\sim$8-9 kcal mol$^{-1}$ from the bimolecular \ce{OH + CO} asymptote (38-39 kcal mol$^{-1}$ from the HOCO potential well) \citep{Ma2012, masunov_catalytic_2018}. Moreover, we could not find this intermediate in the simulations on the \ce{H2O} cluster. We, therefore, skip the search for this intermediate in all cluster calculations. Nonetheless, we discuss the origin of this disagreement in \Cref{sec:appendix1}.

\subsection{Kinetic Analysis}

We employed the microcanonical flavour of the transition state theory, called Rice–Ramsperger–Kassel–Marcus (RRKM) to compute the energy-dependent rate constants $k(E)$ for the transitions between reaction wells, given by:

% \begin{equation}
%     k(E) = \dfrac{\kappa }{h} \dfrac{N^{\#}(E - E_{0})}{\rho(E)},
% \end{equation}

\begin{equation} \label{eq:rrkm}
    k(E) =  \dfrac{N^{\ddagger}(E - E_{0})}{h\rho(E)},
\end{equation}

\noindent where $h$ is the Planck's constant, $N^{\ddagger}(E - E_{0})$ is the sum of states of the transition state evaluated at energy $E$ to the energy of the transition state, $E_0$, and $\rho(E)$ is the density of states of the reactant at energy $E$. In addition, the sum of states contains tunnelling corrections, for which the non-symmetric Eckart potential model was employed \citep{eckart1930penetration, johnston1962tunnelling}.
We did not include rotational symmetry factors in our calculations due to the symmetry breaking induced by the amorphous surface. The rigid-rotor harmonic oscillator model is used throughout the kinetic calculations. The application of RRKM to interstellar reactions is discussed in \citet{Rimola2018} and used or implied in several other works \citep{enrique-romero_theoretical_2021, Perrero2022, baianoGlidingIceSearch2022}

As it will be explained later on (\Cref{sec:rrkm}), the title reaction occurs strictly non-thermally at 10 K. Hence we make our analysis based on $k(E)$ for the entrance \ce{CO + OH -> t-HOCO/c-HOCO} and exit channels: \ce{c-HOCO -> CO2 + H} (and alternatively \ce{c-HOCO/t-HOCO + CO -> CO2 + HCO}, \Cref{sec:pesonCO}). We provide $k(E)$ considering several energy dissipation scenarios. Each of them has a different number of molecules, $n$, over which instantaneous energy dissipation is allowed. We studied $n$=16, 10, 5, and 0 (CO/\ce{H2O}) molecules. In the latter ($n$=0), energy redistribution occurs only within the CO + OH system. We carried out this study by projecting out the molecular Hessian matrix elements for the $m$ molecules (where $m$ = 33 - $n$) farther from the \ce{t-HOCO} minima, as the global minima of our study. The microcanonical rate constants obtained in this study are calculated with the MESS code \citep{Georgievskii2013}. We note that the sizes of the clusters (see Figure \ref{fig:clusters}) and the highest number of dissipating water molecules are sufficient according to previous studies, e.g., \cite{Pantaleone2020, Pantaleone2021}. Although no specific studies have addressed this issue for CO ice, we have made a reasonable assumption that the same holds true. It is worth highlighting again that we considered different dissipating CO ice molecules.

\section{Results} \label{sec:results}

\subsection{Cluster model} \label{sec:res:ices}

\begin{figure*}[t]
    \centering
    \includegraphics[width=0.85\linewidth]{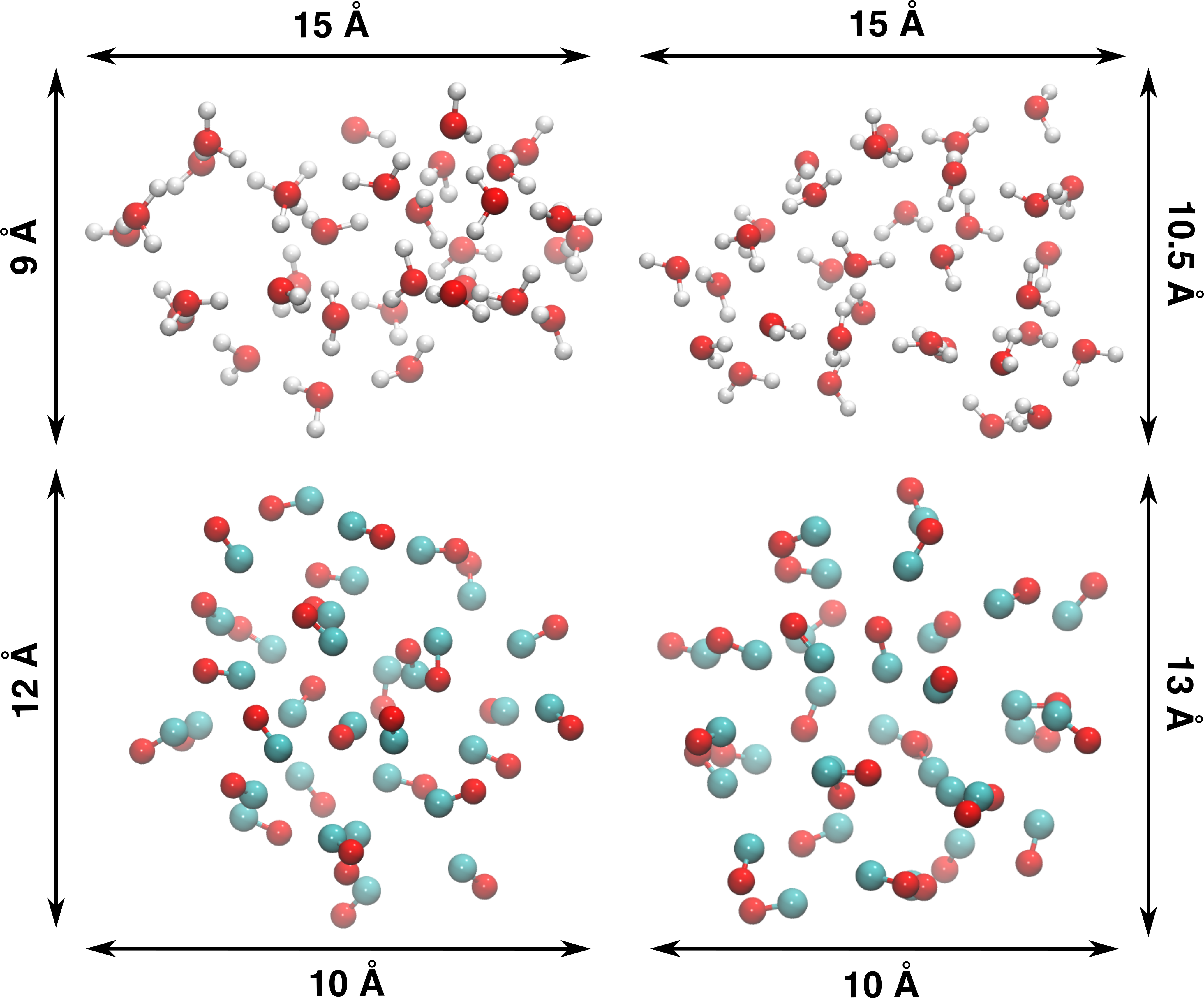}   
    \caption{Views of the clusters employed to simulate the \ce{CO + OH -> CO2 + H} reaction. \emph{Top Left}- Side view of the \ce{H2O} ice cluster. \emph{Top Right}- Top view of the \ce{H2O} ice cluster. \emph{Bottom Left}- Side view of the \ce{CO} ice cluster. \emph{Bottom Right}- Top view of the \ce{CO} ice cluster. Color code: White-Hydrogen, Red-Oxygen, Teal-Carbon. The size of the clusters is determined approximately by measuring the cluster's endpoints. }
    \label{fig:clusters}
\end{figure*}

The fully optimized \ce{H2O} and CO clusters mimicking ice surfaces are presented in \Cref{fig:clusters}. While the CO ice model has a more spherical and compact shape with dimensions 10$\times$12$\times$13~\AA{}, the water one is slightly more elongated, 15$\times$9$\times$10.5~\AA{}. The latter hosts a cavity, where the \ce{CO + OH -> CO2 + H} reaction is simulated. On the contrary, the more compact CO cluster does not have any clear deeper binding site. Hence the reaction site was randomly chosen.

\begin{table}[t]
\begin{center}
\caption{Binding energies (ZPVE corrected, $\Delta$U, in kcal mol$^{-1}$ and Kelvin, in parenthesis), of the CO and OH admolecules at the reaction site considered in this work. The binding energies of t-HOCO and c-HOCO, discussed in \Cref{sec:pes} are also included.}
\label{tab:be}
\begin{tabular}{cc}
\toprule
Adsorbate/Surface & $\Delta$U \\
\bottomrule
CO/\ce{H2O}  & 4.64 (2335) \\
OH/\ce{H2O}  & 6.45 (3246) \\
CO/CO &  0.73 (367) \\
OH/CO &  2.25 (1132) \\
t-HOCO/\ce{H2O}  & 14.51 (7302) \\
c-HOCO/\ce{H2O}  & 12.30 (6190) \\
t-HOCO/\ce{CO}  & 3.52 (1771) \\
c-HOCO/\ce{CO}  & 2.50 (1258) \\
\bottomrule
\end{tabular}
\end{center}
\end{table}

The binding energies of the reactants and reaction intermediates on the surfaces are presented in \Cref{tab:be}. These were calculated as the energy difference between the complexes containing the surface and the admolecule and the sum of the isolated fragments, including ZPVE. In the \ce{H2O} cluster cavity, we find a binding energy for CO  of 4.64 kcal mol$^{-1}$, higher than the values reported by \citet{Ferrero2020} ($\leq$3.71 kcal mol$^{-1}$). This indicates that our cavity is a really deep binding site with a maximized number of neighbour water molecules. For the OH radical, on the contrary, the cavity binding site yields lower than average binding energies (6.45 kcal mol$^{-1}$) than other reported values, e.g., 10.33 kcal mol$^{-1}$ \citep{Duflot2021}, and 10.6 kcal mol$^{-1}$ \citep{enrique2022quantum}. The observed differences arise from the specific structure of our cavity, where the number of dangling H-bonds is saturated, and the binding mode of OH, whose acceptor/donnor H-bonds about 0.1 \AA{} shorter than in the cavity case reported by \citet{enrique2022quantum}. On the CO cluster, the CO/CO binding energy corresponds to the lower bound of the values presented in \citet{Ferrari2023} while the values of OH/CO are unreported. We note that the dual-level error introduced by our calculations is relevant for determining binding energies for CO/CO due to the mismatch of geometries arising from the weak CO-CO interaction in the ice \citep{Ferrari2023}. In the subsequent reactivity studies, the relative magnitude of this error is diminished because energy differences between reaction steps are much higher than the CO-CO interaction energy.

For the reactivity studies, we keep the CO binding site determined above, while the OH radical is placed on a different binding site. We justify this choice based on two arguments. First, when both adsorbates are thermalized, the higher interstellar abundance of CO makes it more likely to be located in deep binding sites, such as the cavity formed in the \ce{H2O} cluster. Second, in \Cref{sec:rrkm}, we investigate the effect of a translationally excited OH radical colliding with a pre-adsorbed CO. 

\subsection{Potential energy surface construction} \label{sec:pes}

All the energy diagrams have been referenced from the asymptotes, i.e., from the sum of energies of the surface, reacting CO and the reacting OH radical.  We will refer to this as the bimolecular system, and for the sake of simplicity it will be denoted as CO + OH, regardless of the ice surface. This was done for the sake of clarity, as it is much clearer what the influence of the substrate in stabilizing the reactants is, as well as its catalytic effect on the barriers.

\subsubsection{\ce{H2O} ice}

We include two pre-reactant complexes following the literature \citep{Ma2012, masunov_catalytic_2018}. First, a pre-reactant complex with large dihedral $\angle \text{HOCO}$ angles, PRC, which leads to the formation of the t-HOCO intermediate. Second, a near 0° dihedral angle pre-reactant complex (PRC'), that forms the c-HOCO intermediate (which was not found on CO ice, as discussed in \Cref{sec:pesonCO}). The transition states that connect the PRCs with the reaction wells are named TS1 and TS1', respectively, and the transition state connecting these two wells is TS2. Finally, the transition state leading to \ce{CO2 + H} from c-HOCO is named TS4. The reason for not naming it TS3 is that the TS3 label (specifically TS3') is reserved for the exit transition state from t-HOCO, a stationary point we do not find on water ice.

\begin{figure}
    \centering
    \includegraphics[width=0.95\columnwidth]{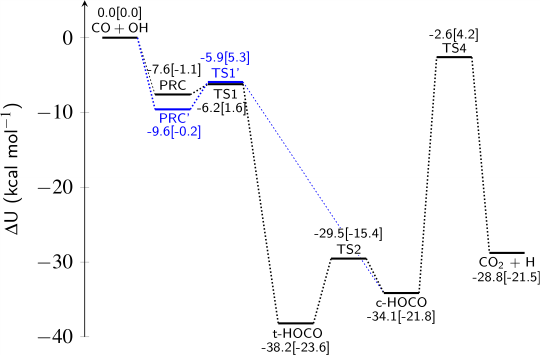}
    \caption{Reaction profile for the \ce{CO + OH -> CO2 + H} reaction on the ASW cluster. Energies are referred to as the sum of the isolated components, e.g. CO, OH, \ce{(H2O)33}. In square brackets, the energies for the gas-phase reaction. All energies are ZPVE corrected.}
    \label{fig:pesh2o}
\end{figure}

\begin{figure}
    \centering
    \includegraphics[width=1.00\linewidth]{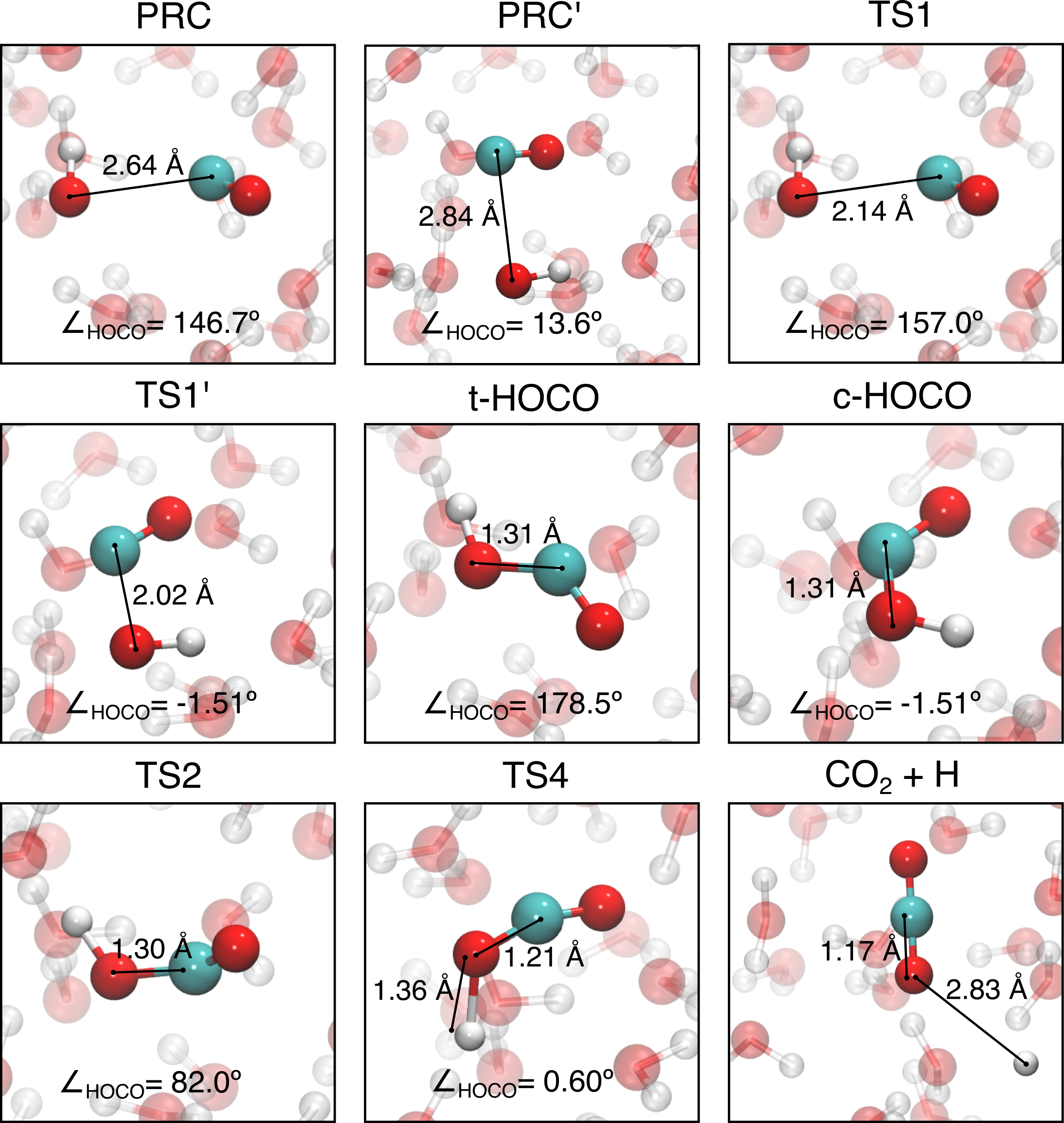}
    \caption{Stationary points on the \ce{CO + OH -> CO2 + H} reaction profile for the reaction on top of the \ce{(H2O)33} cluster. For all the intermediates and transition states we indicate the value of the dihedral angle $\angle$HOCO governing the isomerism of the t-HOCO and c-HOCO wells. Color code. White-Hydrogen, Red-Oxygen, Teal-Carbon.  }
    \label{fig:strh2o}
\end{figure}

The stationary points on the reaction profile are gathered in \Cref{fig:pesh2o}. The reaction profile has, for the most part, the same profile as in the gas phase, with two notable exceptions. The first concerns the absence of the \ce{HCO2} intermediate, as we already discussed in \Cref{sec:quantum}. The second is the inversion in energy between PRC and PRC'. This inversion appears following the formation of a \ce{HO--H2O} hydrogen bond that locks the PRC' geometry in the binding site contiguous to the CO binding site. The snapshots of the stationary points are collated in \Cref{fig:strh2o}, where this effect can be visualized. The higher stabilization of PRC' also results in higher activation energy to c-HOCO through TS1'.  

The binding energies of t-HOCO  and c-HOCO on the cavity are 15.51 kcal mol$^{-1}$ (7805 K) and 12.30 kcal mol$^{-1}$ (6190 K), respectively. These binding energies are significantly higher than the ones for CO and OH presented in \Cref{tab:be}, and are closer to the average values reported for the related molecule, HC(O)OH, formic acid (e.g., $\sim$ 12.30 kcal mol$^{-1}$ \cite{Molpeceres2022Thio}, 10.7--21.0 kcal mol$^{-1}$ \cite{Ferrero2020}). The t-HOCO and c-HOCO wells are significantly stabilized on the surface, evinced by the 13--16 kcal mol$^{-1}$ difference in energy with the same intermediates in the gas phase. As a consequence, the activation energy of TS4 is higher on water.
When breaking the O--H bond in c-HOCO, the energy corresponding to the OH moiety must be overcome, i.e. a significant fraction of the binding energy. The binding energy of the \ce{CO2 + H} system on \ce{H2O} was found to be 7.30 kcal mol$^{-1}$ (3673 K).

Finally, from \Cref{fig:pesh2o}, it is evident that the reaction, if viable, must proceed through quantum tunnelling. The \ce{c-HOCO -> CO2 + H} barrier is 32.1 kcal mol$^{-1}$, which is extremely high for ISM conditions. However, contrary to what happens in the gas phase, TS4 is submerged with respect to the reactant asymptote, thanks to the stabilization promoted by the \ce{H2O} surface. The product of the reaction, \ce{CO2 + H}, is higher in energy than both radicals, and the reaction is significantly less exothermic because of the break of hydrogen bonds. Nonetheless, once \ce{CO2 + H} is formed, H is susceptible of diffusing or evaporating, thus concluding the reaction.

\subsubsection{\ce{CO} ice} \label{sec:pesonCO}

\begin{figure}
    \centering
    \includegraphics[width=0.95\columnwidth]{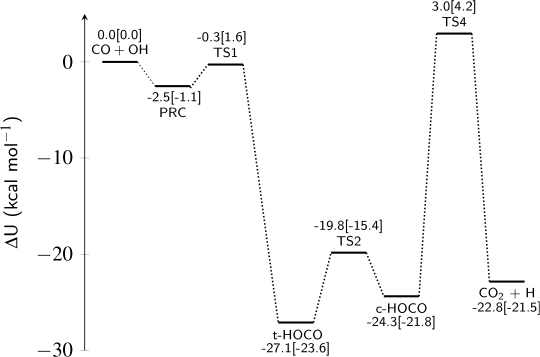}
    \caption{Reaction profile for the \ce{CO + OH -> CO2 + H} reaction on the amorphous CO cluster. Energies are referred to as the sum of the isolated components, e.g. CO, OH, \ce{(CO)33}. In square brackets, the energies for the gas-phase reaction. All energies are ZPVE corrected.}
    \label{fig:pesco}
\end{figure}

\begin{figure}
    \centering
    \includegraphics[width=1.00\linewidth]{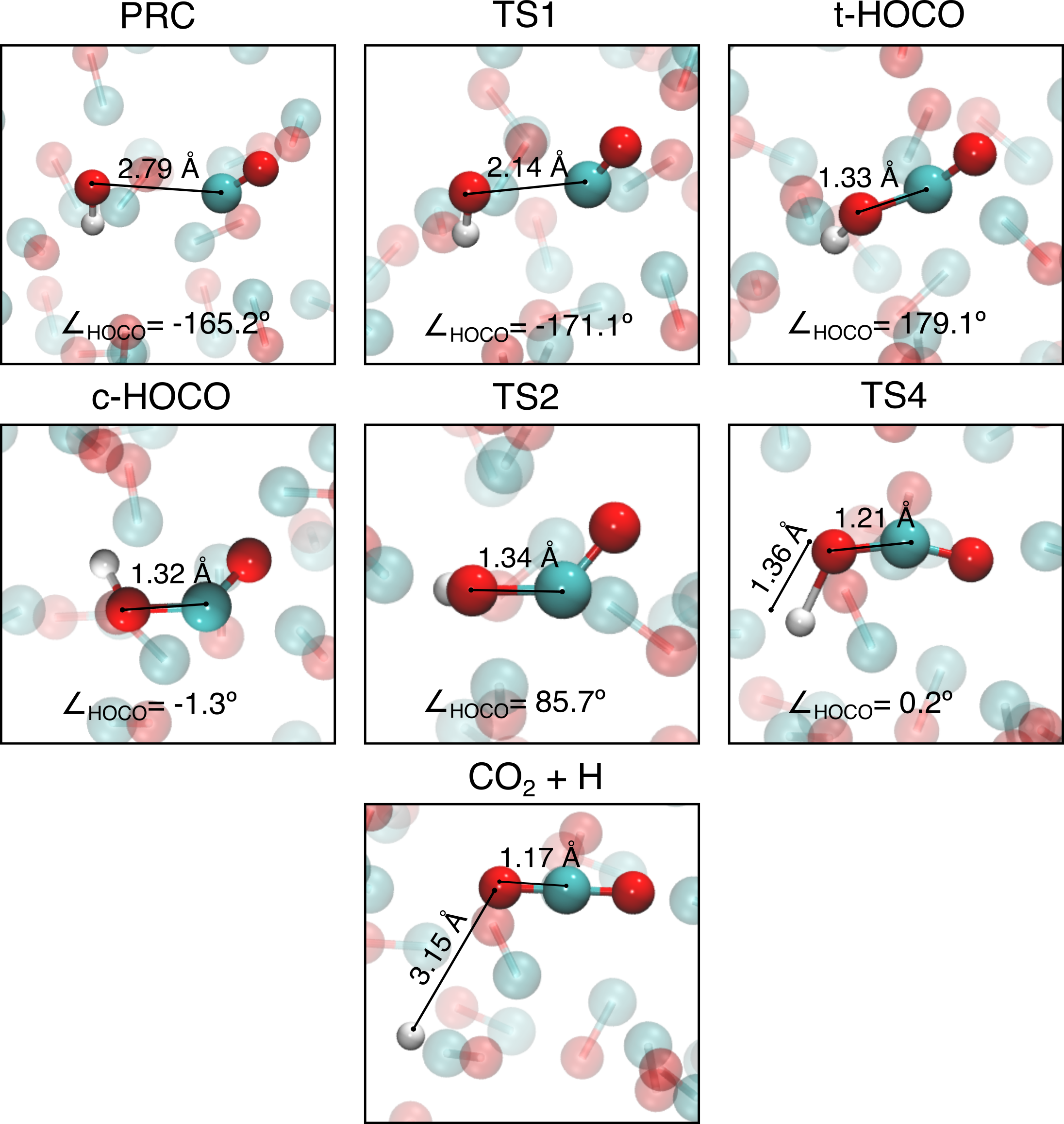}
    \caption{Stationary points on the \ce{CO + OH -> CO2 + H} reaction profile for the reaction on top of the \ce{(CO)33} cluster. For all the intermediates and transition states we indicate the value of the dihedral angle $\angle$HOCO governing the isomerism of the t-HOCO and c-HOCO wells. Color code. White-Hydrogen, Red-Oxygen, Teal-Carbon.  }
    \label{fig:strco}
\end{figure}

The reaction profile on CO ice is shown in Figure \ref{fig:pesco} and the stationary points in Figure \ref{fig:strco}. With respect to the gas-phase process, as previously discussed, the profile lacks the \ce{HCO2} intermediate. When comparing with the results for the water cluster presented above, the main difference is the lack of PRC', so that the reaction must go through the \ce{t-HOCO} intermediate to reach \ce{CO2}. While PRC' exists on the CO ice, we found it to be a first-order saddle point. Unlike in water, where PRC' is stabilized thanks to the interaction of the OH radical with a dangling bond of \ce{H2O}, on CO, this interaction is unavailable, and the weak OH-CO interaction promotes the rotation to PRC. There is still the possibility that the lack of PRC' is an effect of the random selection of the binding site, however a full binding site sampling is beyond our computational resources.
To reach the \ce{t-HOCO} intermediate, however, the TS1 must be crossed at the same energy level as the asymptote. Hence, significant energy dissipation would suppress the whole reaction unless enough energy input is provided via non-thermal mechanisms. 

Additionally, the much reduced inter-molecular interaction of the admolecules with the surface due to the lack of electrostatic and H-bonding interactions of CO ices affects the energetics of the stationary points. The most prominent examples are the lower stabilisation of intermediates and the barrier in TS4, which sits above the energy of the asymptote. In general, the energetics on CO ice is closer to the gas phase case, with small differences, e.g., the isomerisation barrier for the \ce{t-HOCO -> cis-HOCO} reaction on CO is about 1 kcal mol$^{-1}$ lower (and about 2 kcal mol$^{-1}$ lower for the reverse reaction).

\begin{figure}
    \centering
    \includegraphics[width=\linewidth]{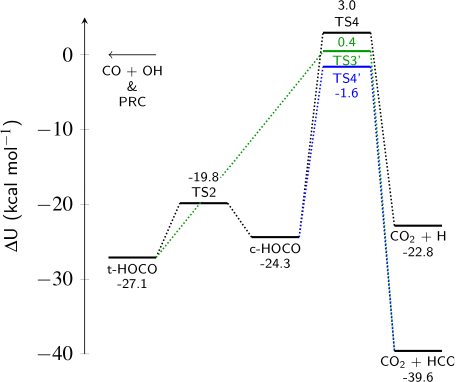}
    \caption{Reaction profile for the \ce{CO + OH -> CO2 + H} and \ce{2CO + OH -> CO2 + HCO} alternative reaction on the amorphous CO cluster. Energies are referred to as the sum of the isolated components, e.g. CO, OH, \ce{(CO)33}. Note that we omit the separated reactants asymptote and prc energies, already presented in \Cref{fig:pesco}. All energies are ZPVE corrected.}
    \label{fig:pescoalt}
\end{figure}

\begin{figure}
    \centering
    \includegraphics[width=\linewidth]{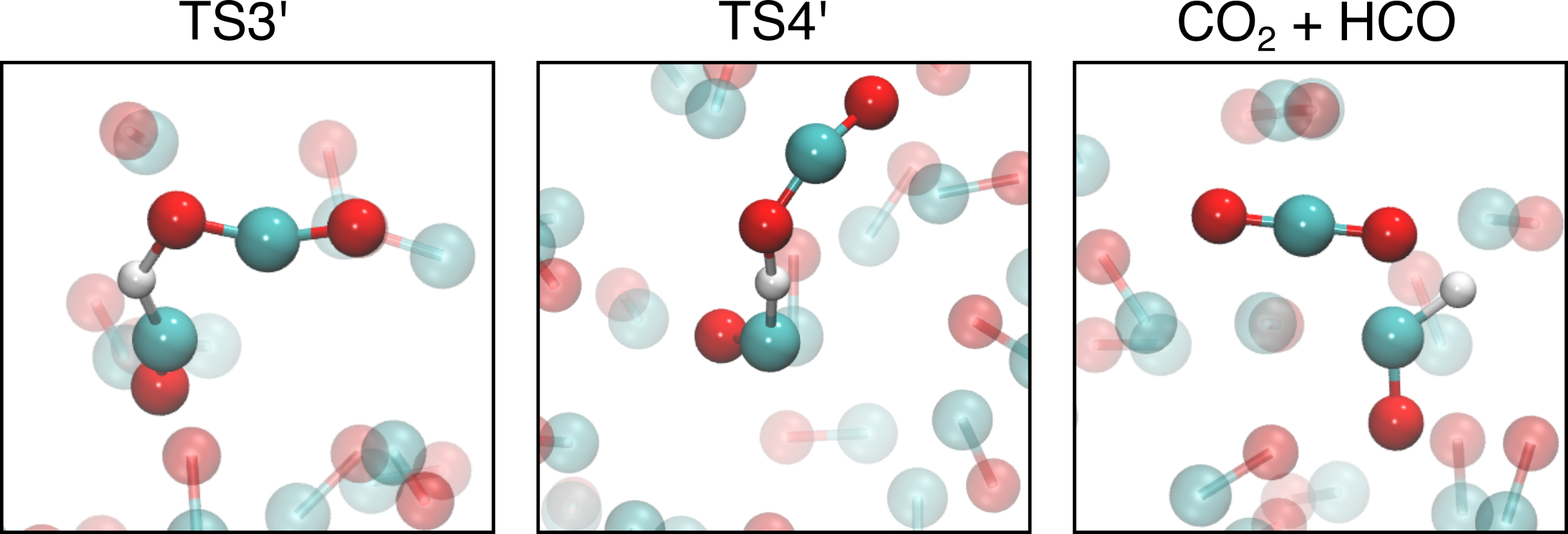}
    \caption{Additional stationary points to the ones presented in \Cref{fig:strco} for the alternative reaction channel \ce{2CO + OH -> CO2 + HCO}. Color code. White-Hydrogen, Red-Oxygen, Teal-Carbon.  }
    \label{fig:strcoal}
\end{figure}

The fact that there are more CO molecules surrounding the reaction site opens a new possibility not available on water ice or the gas phase. It involves the reactivity of the \ce{t-HOCO} and \ce{cis-HOCO} intermediates with a neighbouring CO, leading to \ce{CO2} + HCO, see Figure \ref{fig:strcoal}. Interestingly, these reactions possess lower activation energy barriers than TS4, see Figure \ref{fig:pescoalt}, and in the case of the \ce{cis-HOCO + CO ->  CO2 + HCO} reaction, the barrier sits below the asymptote.

\subsection{Microcanonical rate constants} \label{sec:rrkm}

We estimated the microcanonical rate constants for the PES entrance and exit channels described in the previous sections. The entrance channels start with the pre-reactant complexes and finish with t/c-HOCO, and the exit channels start with t/c-HOCO and finish with \ce{CO2 + H}, and additionally \ce{CO2 + HCO} for CO. These channels present the relevant rate constants for the kinetics of the reaction because the \ce{t-HOCO -> c-HOCO} is much faster, even when energy redistribution is at play. Notice that due to the barriers (TS1 and TS1'), if the stationary points of the PES were populated according to a thermal distribution, the formation of the HOCO intermediates would be slow, and the formation of products would likely not happen at all. To simulate non-thermal reactions, an initial amount of energy is given to the system; see below.
Experiments of \citet{Oba2010Carbonic} show the formation of HOCO with an apparent small barrier or null barrier.  We note that for the exit channel \ce{(c/t)-HOCO -> CO2 + H/HCO }, the starting potential well is very deep, and thermalization is more likely \citep{arasa_molecular_2013}. Nevertheless, as we will show, under a microcanonical formalism, the formation of \ce{CO2 + H} is found to be slow. Finally, different energy dissipation is allowed by changing the number of ice molecules considered in the microcanonical calculations, $n$.

Our PESs indicate that adsorption energy (formation of PRC/PRC') is not completely dissipated but employed in forming HOCO. The energy reference is again the energy of the asymptotes. One could consider that this is not the best choice since the initial energy lies above the energy of the PRC/PRC' and it would actually mean that the initial state is actually higher in energy than a fully thermalized reactant set. However, it must be noted that (i) if a reference state is an upper bound of the real one, and even in this case the reaction is not plausible, then starting from a more stable reference will not change the qualitative picture, and (ii) in cases where an incomplete energy dissipation promoted by certain exothermic processes, e.g. diffusion into deeper binding sites and possible Eley-Rideal mechanisms \footnote{That may be of relevance for CO molecules given their abundance in ISM ices.} would actually involve higher initial energies than PRC/PRC'. This effect is irrelevant when the activation energy of a reaction is much higher than the exothermicity caused by the mentioned processes, but for \ce{CO + OH -> HOCO} the activation energy of the reaction falls below the adsorption energy, and it is of small magnitude. The correct energy reference would lie somewhere in between that of the asymptote and the PRC/PRC'.

\begin{figure}
    \centering
    \includegraphics[width=1.00\linewidth]{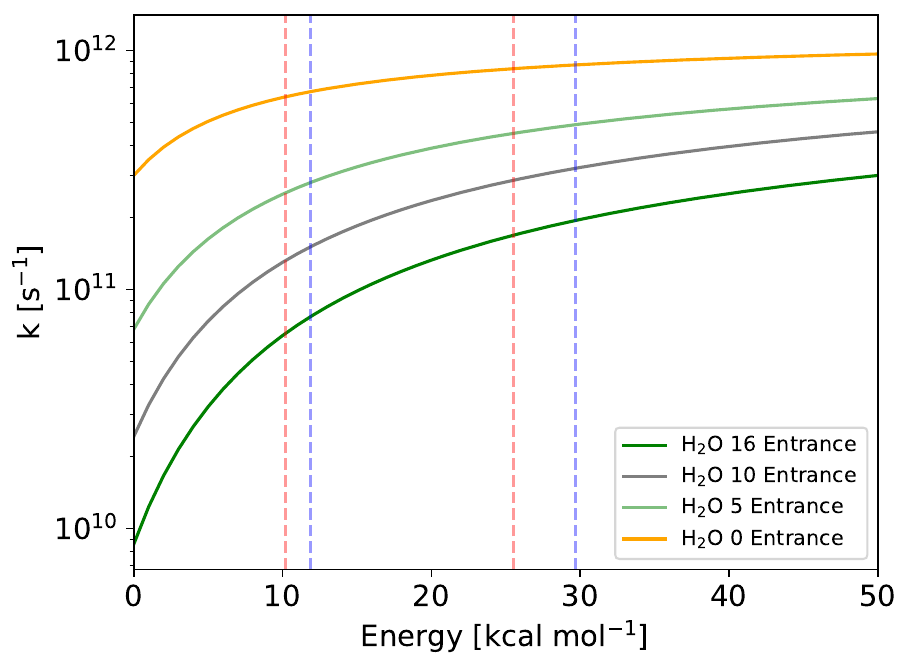} \\
    \includegraphics[width=1.00\linewidth]{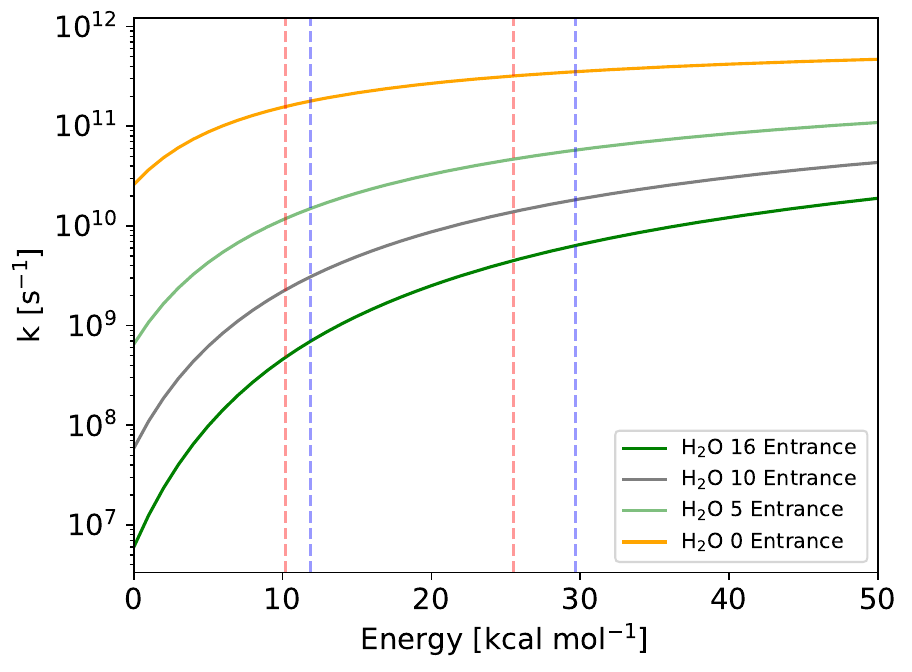} \\
    \caption{Microcanonical (energy-dependent) rate constants for the \ce{CO + OH -> t-HOCO} (top) and \ce{CO + OH -> c-HOCO} (bottom) reaction step on \ce{H2O} ice. The different solid lines represent different ergodic energy dissipation scenarios, with $n$ molecules accepting the reaction energy. The vertical dashed lines indicate fractions (0.1 and 0.25) of the \ce{O + H} reaction (red) and of half the energy of a Lymann alpha photon (blue); see text. The zero of energy is defined as the PES asymptote (e.g. no adsorption). }
    \label{fig:entranceh2o}
\end{figure}

\begin{figure}
    \centering
    \includegraphics[width=1.00\linewidth]{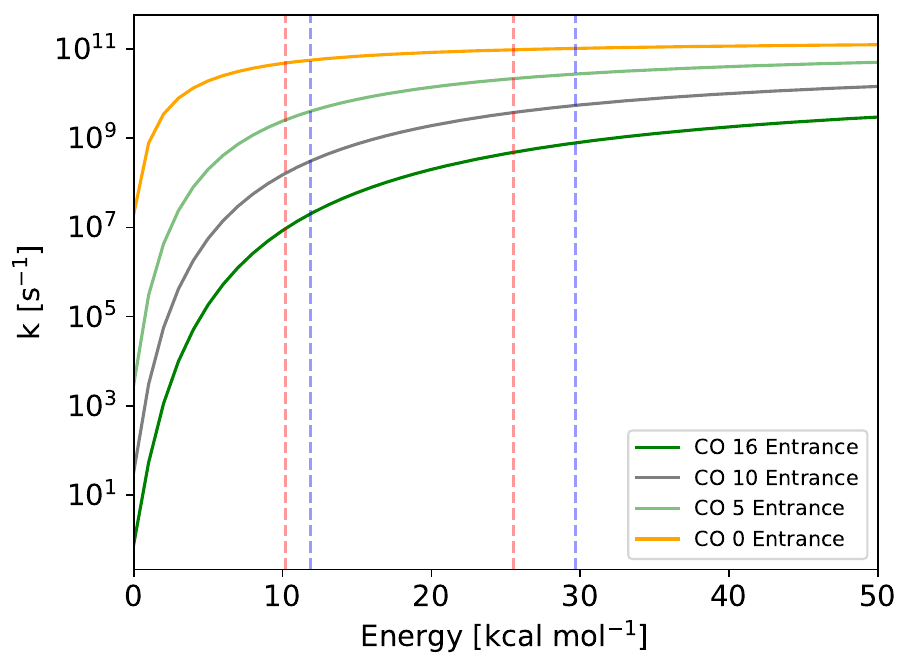}
    \caption{Microcanonical (energy-dependent) rate constants for the \ce{CO + OH -> t-HOCO} reaction step on CO ice. The different solid lines represent different ergodic energy dissipation scenarios, with $n$ molecules accepting the reaction energy. The vertical dashed lines indicate fractions (0.1 and 0.25) of the \ce{O + H} reaction (red) and of half the energy of a Lymann alpha photon (blue); see text. The zero of energy is defined as the PES asymptote (e.g. no adsorption). }
    \label{fig:entranceco}
\end{figure}

\begin{figure}
    \centering
    \includegraphics[width=1.00\linewidth]{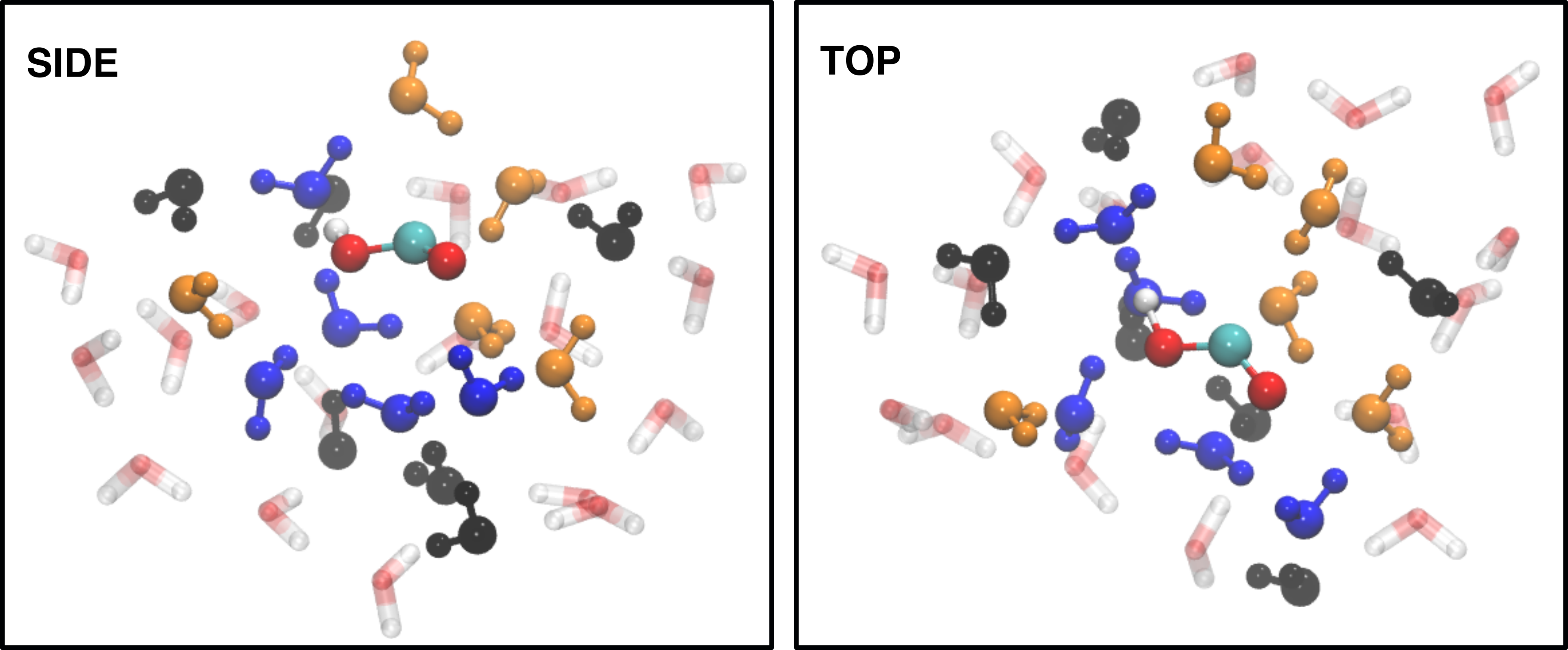}
    \caption{Graphic representation of the different layers admitting energy dissipation through RRKM equipartition in the \ce{H2O} model ice. The t-HOCO molecule (0 dissipating molecules) is represented with red, teal and white spheres. the 5 \ce{H2O} model includes the t-HOCO molecule and \ce{H2O} molecules portrayed in blue. The 10 and 16-molecule model are represented with gold and black molecules, respectively. Finally, the rest of the molecules are represented as transparent liquorices.}
    \label{fig:dissh2o}
\end{figure}

\begin{figure}
    \centering
    \includegraphics[width=1.00\linewidth]{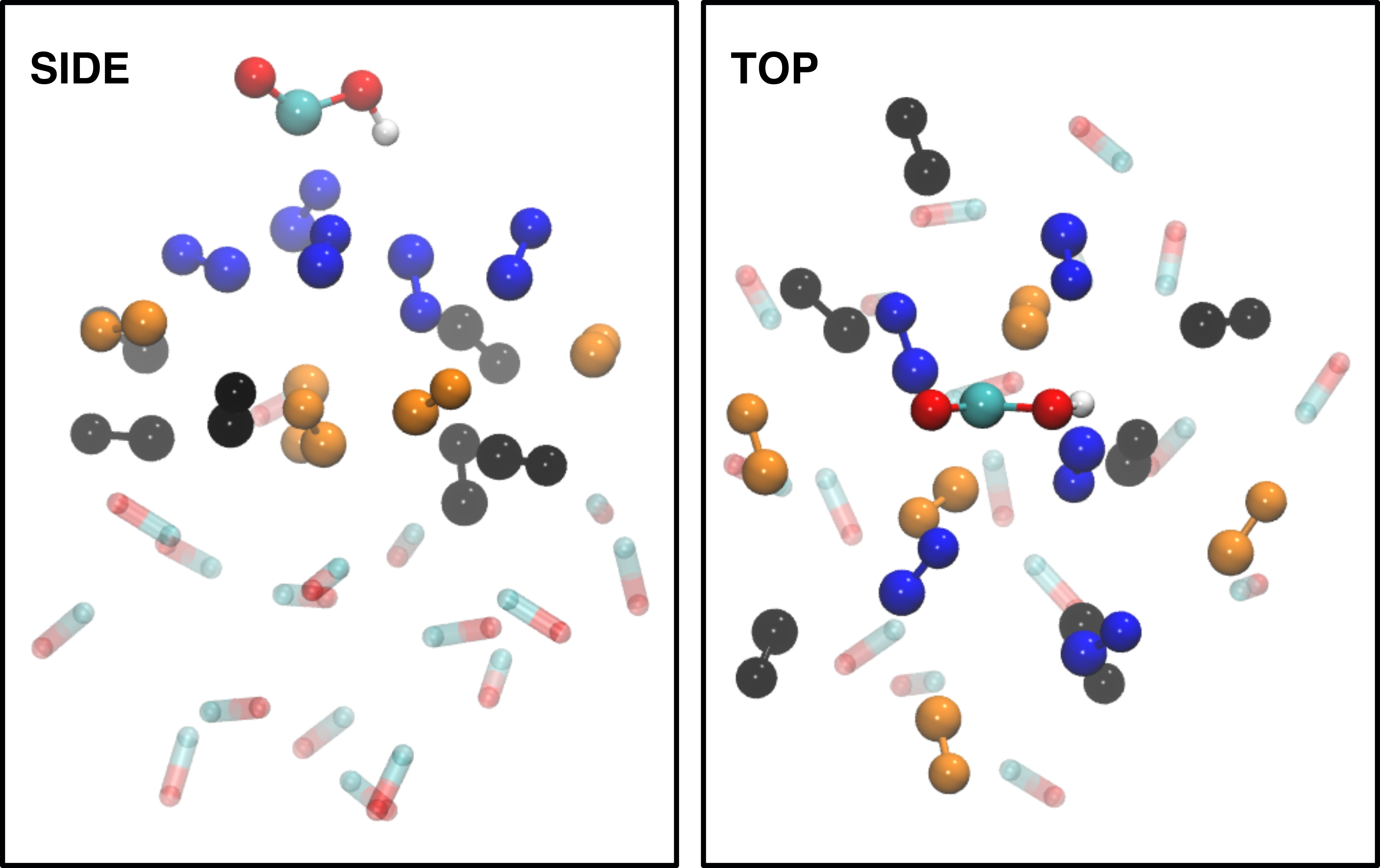}
    \caption{Graphic representation of the different layers admitting energy dissipation through RRKM equipartition in the CO model ice. The t-HOCO molecule (0 dissipating molecules) is represented with red, teal and white spheres. The 5 CO model includes the t-HOCO molecule and CO molecules portrayed in blue. The 10 and 16-molecule model are represented with gold and black molecules, respectively. Finally, the rest of the molecules are represented as transparent liquorices.}
    \label{fig:dissco}
\end{figure}

The microcanonical rate constants for the entrance step are shown in \Cref{fig:entranceh2o} and \Cref{fig:entranceco} for \ce{H2O} and \ce{CO} ice. In this plot, we show the reaction rate constants as a function of the energy, where $k(E=0)$ corresponds to the separated, no adsorption asymptote (CO + OH in \Cref{fig:pesh2o} and \Cref{fig:pesco}). Energies above zero indicate extra energy from non-thermal excitation mechanisms.
% Above zero, the input energy in the ISM can come from various non-thermal factors.
In this work, to compare with experimental observations, we will consider the presence of extra energy from either (i) a prior \ce{O + H -> OH} reaction ($\Delta$U = 102.1 kcal mol$^{-1}$) or (ii) half the energy deposited by a single Ly-$\alpha$ photon, assuming equal energy partition into the products of the \ce{H2O -> OH + H}, ($\Delta$U = 118.7 kcal mol$^{-1}$). Notice that the amount of extra energy used to promote the title reaction through the non-thermal mechanisms is unknown. Hence, we represent fractions of that energy, 0.10, 0.25, 0.50, as vertical dashed lines in \Cref{fig:entranceh2o} and \Cref{fig:entranceco} to serve as a guide to evaluate how the rate constants would increase under these assumed scenarios. As we introduced in \Cref{sec:methods}, we evaluated the behaviour of the reaction assuming dissipation into a set of $n$ molecules. The four different cases for $n$=0, 5, 10, 16 are illustrated in \Cref{fig:dissh2o} and \Cref{fig:dissco}.

The rate constants for the entrance step on \ce{H2O} ice are, for all $n$ dissipating molecules, fast for the \ce{PRC -> t-HOCO} step, indicating that external energy input is unnecessary for this reaction, as determined experimentally by \citet{Oba2010Carbonic} and computationally by \citet{arasa_molecular_2013}. However, for the alternative PRC' $\rightarrow$ c-HOCO reaction, we observe $k(E=0)\leq10^{8}$ s$^{-1}$ for the models with 10, 16 \ce{H2O} dissipating molecules. This means that if the timescale for thermalization is shorter than tens of nanoseconds, the adsorption energy alone is insufficient to overcome the entrance barrier. This constraint is lifted by considering extra energy. The reason for the difference between rate constants for the reactions starting from PRC and PRC' stems from the significantly higher activation energy in the latter case.

For the CO model, we observe systematically lower values of $k(0)$ than in water, owing to the lower stabilization of the PRC complex on CO than on \ce{H2O} leading to higher energy barriers than in the best case for \ce{H2O}. This, in turn, yields $k(E=0)\leq10^{8}$ s$^{-1}$ for all of our models. Because $k(E)$ is a very steep function around $E$=0, the reaction is viable with a small input of energy that can come from reactions, e.g. \ce{O2 + H} \citep{Qasim2019}. This finding reinforces the scenario presented in \citet{Garrod2011} for the three body formations of \ce{CO2} on CO ice, as we will discuss in \Cref{sec:discussion}. An important comment for each of these rate constants is that we implicitly assumed an infinitely fast energy partition into $n$ molecules, which may not be a good representation of this reaction on CO. At this research stage, we warn that extracting strong conclusions for a limit case like the one found for \ce{PRC -> t-HOCO} on CO ice is difficult and more sophisticated approaches are necessary. We are currently working on a molecular dynamics study of this reaction to illuminate this issue.

\begin{figure}
    \centering
    \includegraphics[width=1.00\linewidth]{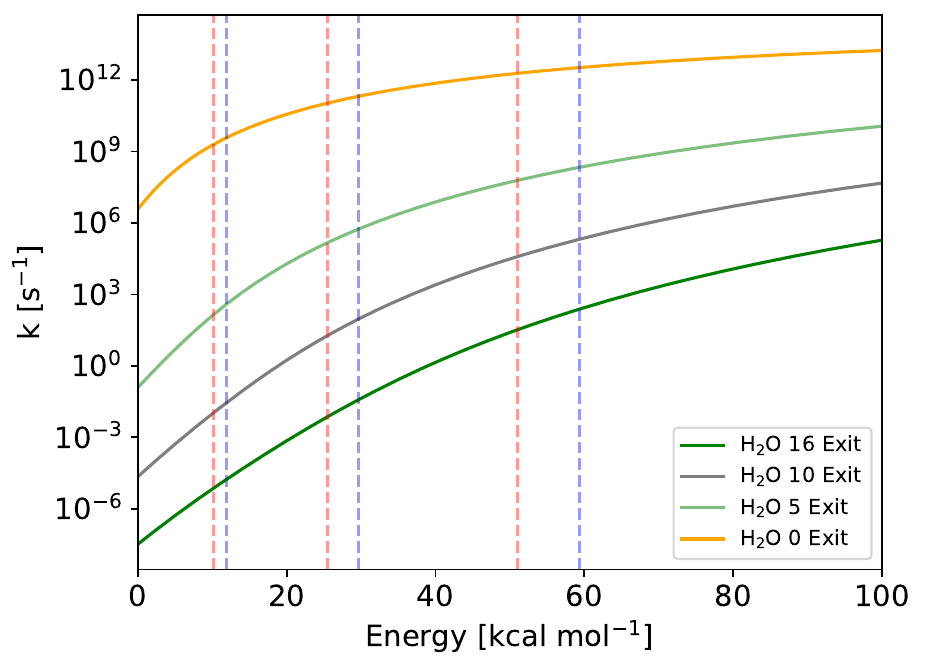} \\
    \caption{Microcanonical (energy-dependent) rate constants for the \ce{c-HOCO -> CO2 + H} reaction step on \ce{H2O} ice. The different solid lines represent different ergodic energy dissipation scenarios, with $n$ molecules accepting the reaction energy. The vertical dashed lines indicate fractions (0.1, 0.25 and 0.5) of the \ce{O + H} reaction (red) and of half the energy of a Lymann alpha photon (blue); see text. The zero of energy is defined as the PES asymptote (e.g. no adsorption).}
    \label{fig:exitH2O}
\end{figure}

\begin{figure}
    \centering
    \includegraphics[width=1.00\linewidth]{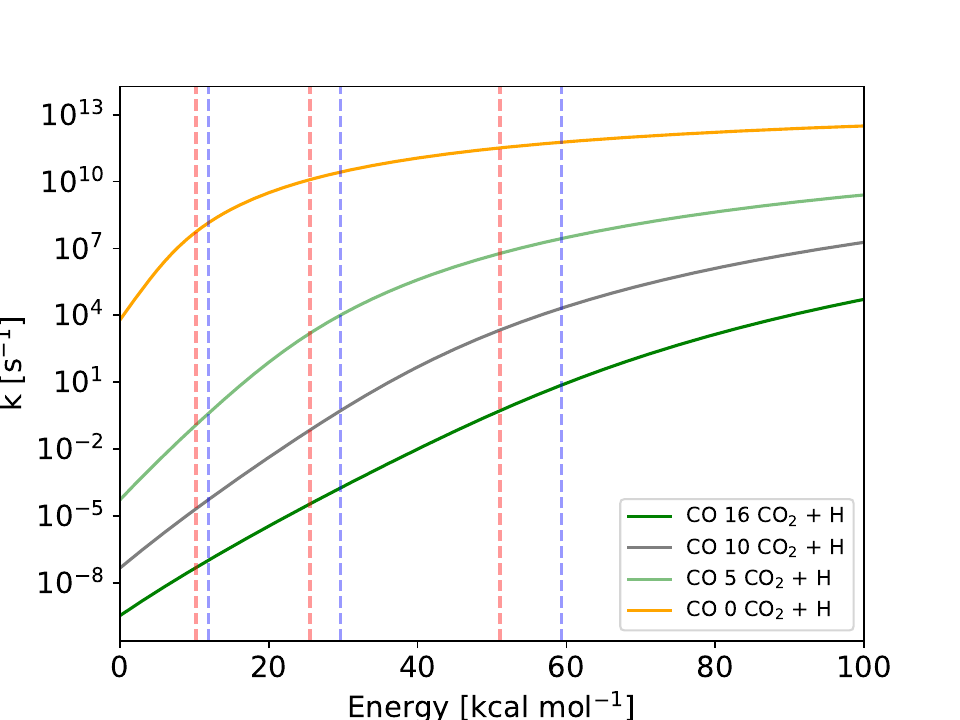} \\
    \includegraphics[width=1.00\linewidth]{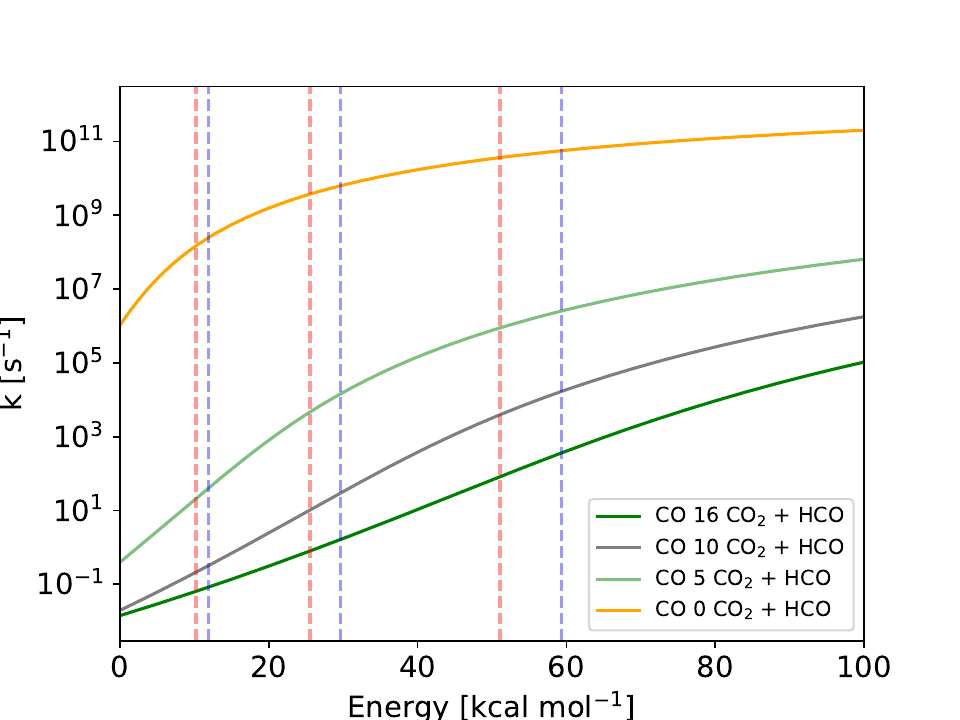} \\
    \caption{Microcanonical (energy-dependent) rate constants for the \ce{c-HOCO -> CO2 + H} (top) and \ce{c-HOCO + CO -> CO2 + HCO} (bottom) reaction step on \ce{CO} ice. The different solid lines represent different ergodic energy dissipation scenarios, with $n$ molecules accepting the reaction energy. The vertical dashed lines indicate fractions (0.1, 0.25 and 0.5) of the \ce{O + H} reaction (red) and of half the energy of a Lymann alpha photon (blue); see text. The zero of energy is defined as the PES asymptote (e.g. no adsorption).}
    \label{fig:exitco}
\end{figure}

Similarly to the entrance rate constants, the exit \ce{c-HOCO -> CO2 + H} rate constants on \ce{H2O} ice and \ce{c/t-HOCO -> CO2 + H/HCO} rate constants on CO ice are plotted in  \Cref{fig:exitH2O} and \Cref{fig:exitco} for the different dissipation scenarios. It is important to remind that while the entrance channels are unaffected by quantum tunnelling, all the exit channels involve the migration of an H atom, turning quantum tunnelling into an important driver for the reaction, as already evinced by nuclear quantum dynamics calculations \citep{Ma2012}. Still, even with the influence of quantum tunnelling, the reactions are, in all cases, significantly slower than in the entrance step. The importance of the energy dissipation scheme is major for these reactions. There is a clear gap in exit rate constant values between the (ideal) $n$=0 dissipation model and the 5, 10 and 16 molecules dissipation models that, in all the cases, yield rate constants $k(E=0)\leq$ 0 s$^{-1}$. We remind that these values must be confronted against the thermalization timescale, i.e. if thermalization is faster, the reaction will not proceed. A rate constant of $k(E=0)\leq$ 0 s$^{-1}$ means reaction times of seconds, and we find it hard that thermalization would not happen on those timescales, precluding all the \ce{c/t-HOCO -> CO2 + H/HCO} reactions in all the conditions and substrates considered in this work. We conclude then that, without the input of any external energy other than the adsorption energy of the reactants, the reaction can proceed neither microcanonically nor from thermalized HOCO. 

When including a degree of external energy from the mechanisms explained above (chemical and \ce{H2O} photodissociation), the exit reaction is faster, as expected. However, only the $n$=0 dissipation model yields rate constants that are sufficiently high $\geq$ 10$^{8}$ s$^{-1}$ to compete with thermalization. The upper bound of the timescale for (almost) complete thermalization of HOCO is estimated to be similar to that of \ce{CO2} formed from the \ce{CO + ($^1$D)O -> CO2} reaction, that is, a few nanoseconds \citep{Upadhyay2021}. While the energy dissipation in RRKM is instantaneous, and an incomplete energy dissipation may increase the values of the rate constants, our assumption for the external energy input is also rather ideal. Thus, we conclude that even in the presence of external energy input, we find it hard to justify the formation of \ce{CO2} and \ce{H/HCO} from the title reaction. This suggests that the formation of \ce{CO2} relies on the subsequent reaction described as follows:

\begin{align}
    \ce{t/c-HOCO + H &-> CO2 + H2}. \label{eq:abstraction}
\end{align}

Reaction \ref{eq:abstraction} involves two radicals, and even though an activation barrier may be present on ice \citep{enrique2022quantum} quantum tunnelling should play a major role, as it is the case found for H abstraction reactions \citep{Molpeceres2022Thio, molpeceres_processing_2023}. Thus, reaction \ref{eq:abstraction} must be viable. The inclusion of reaction \ref{eq:abstraction} in the \ce{CO2} reaction network was already in place for the non-energetic formation of \ce{CO2}, for example, in \citet{Qasim2019}. Still, this article shows that it also applies to the energetic formation of \ce{CO2}. We put our results in a laboratory/simulation and astrophysical context in \Cref{sec:discussion}.

Finally, and despite it does not affect the outcome of the reactions studied in this work (e.g. the \ce{t/c-HOCO ( + CO) -> CO2 + H/HCO} reactions remain non-viable under ISM conditions), it is interesting from a purely chemical perspective to comment on the effect observed for the two competing reactions \ce{c-HOCO -> CO2 + H} and \ce{t/c-HOCO + CO -> CO2 + HCO}. The competition between these two processes is energy dependent. At low values of $E$, e.g. $k(E=0)$, favours \ce{t/c-HOCO + CO -> CO2 + HCO} whereas \ce{c-HOCO -> CO2 + H} is the preferred exit channel at higher energies, between 10--120 kcal mol$^{-1}$, depending on the number of dissipating molecules. The dependence on the energy and number of dissipating molecules clearly reveals that the dominion of the \ce{c-HOCO -> CO2 + H} route at high energies is an entropic effect. For both routes, the count of states at the TS energy (the numerator of \Cref{eq:rrkm}) depends on the height of the barrier and the number of low-frequency vibrational modes. Because HCO, in contrast with H, has two molecular vibrations, \ce{H-C} and \ce{C=O}, at 2800 and 1900 cm$^{-1}$, the count of states will be smaller at high energies. Low-frequency vibrations overwhelm the purely kinetic effect arising from the lower barrier.

\section{Discussion} \label{sec:discussion}

\subsection{The \ce{CO + OH -> CO2 + H } reaction in the laboratory} 

The experiments carried out in the \ce{CO + OH -> CO2 + H } reaction were reviewed in \Cref{sec:introduction}. For most of them, the biggest experimental conundrum is the generation of the OH radical, which is very unstable under laboratoty conditions and needs to be generated in situ. The experimental methods for forming the OH radical in these experiments are, in most cases, different. However, all the possible formation pathways involve the co-deposition or co-generation of H atoms e.g. formation with \ce{O2 + H}, fragmentation of \ce{H2O} in a microwave discharge or \ce{H2O} photodissociation. In general, it is impossible to experimentally discern whether the \ce{CO + OH} reaction proceeds directly to \ce{CO2 + H} or, in turn, stops at t-HOCO, which is converted to \ce{CO2} via reaction \ref{eq:abstraction}.

A rigorous study of the reaction using molecular dynamics \citep{arasa_molecular_2013} showed the probability of direct formation of \ce{CO2} on \ce{H2O} ice is lower than 1\%. It is important to remark that in \citet{arasa_molecular_2013}, the OH was generated with excess energy coming from photodissociation of \ce{H2O}. Our results support the latter scenario and discard the direct reaction. Compared with our results, the small fraction observed for the direct formation of \ce{CO2 + H} in \citet{arasa_molecular_2013} may come from the slower and more realistic non-ergodic energy dissipation present in the molecular dynamics study. 

On CO ice, the reaction proceeds similarly to in \ce{H2O}, both in our calculations and in the experiments of \citet{Qasim2019}, where HOCO is explicitly included as the intermediate for the reaction. \citet{Qasim2019} discuss the competition with formic acid (HC(O)OH) through the reaction:

\begin{align}
    \ce{HOCO + H &-> HC(O)OH}
    \label{reac:FA}
\end{align}

\noindent with Reaction \ref{eq:abstraction}.
Our results complement these experiments as well, showing that in addition to what was already known, the formation of the HOCO complex has to surmount an activation energy of 2.2 kcal mol$^{-1}$ with a mere adsorption energy of 2.5 kcal mol $^{-1}$, in contrast with \ce{H2O} ice, where the higher stabilization of the PRC complex increases the energetic budget for the formation of HOCO. The consequence of this effect in the overall reaction scheme is that the formation of HOCO cannot be taken for granted on CO ice under a non-energetic regime. In \citet{Qasim2019}, such energy input is given by a preceding chemical reaction. The more impeded formation of the HOCO radical on CO is the main difference with \ce{H2O} ice and is illustrated by the rate constants in \Cref{fig:entranceh2o} (Top panel) and \Cref{fig:entranceco}. This different reactivity on different substrates may explain the recent JWST observations of a higher degree of mixing of \ce{CO2} with \ce{H2O} than with \ce{CO} \citep{McClure2023}. However, and as we indicated in section \Cref{sec:rrkm}, further studies are being undertaken to understand the precise behaviour of the \ce{CO + OH -> t-HOCO} association step on CO ices.

On the other hand, \citet{2021GutierrezQuinanilla} used matrix isolation, electron paramagnetic resonance and FT-IR techniques, which made it possible to observe several radicals, among which HOCO, and \ce{CO2}. \ce{HC(O)OH} is also detected, although its formation seems to be due to HCO + OH rather than reaction \ref{reac:FA}. In this experiment, methanol molecules embedded in an Argon matrix are photolysed at 14 K. The resulting photo-products can relax as the matrix acts as a third body. Later the sample is warmed up to 35 K, and the Ar matrix is removed, allowing light species to diffuse. The peak of CO$_2$ production occurs in this last stage. According to our results and interpretation, if \ce{CO2} is formed via reaction \ref{reac:1}, either there is some extra energy input, not all the energy from the photolysis step was completely dissipated, or H-abstraction reactions are in place. In the latter case, this can be triggered by other radicals rather than reaction \ref{eq:abstraction}, something we did not consider in this work, and that would require either the diffusion at warmer temperatures or the presence of a nearby radical species. In addition, an efficient H-abstraction radical-radical channel should be present, which will certainly depend on their relative orientation \citep{enrique2022quantum}. Notice that in this experiment, no ice surface is present, but rather the bare copper plate on top of which the matrix and reactant mixture is prepared. Finally, we would like to encourage more experiments on CO$_2$ formation starting from thermalized reactants, especially on CO surfaces.

\subsection{The \ce{CO + OH -> CO2 + H } reaction in the ISM} 

The comparison between the experiments and our calculations presented in the last section motivates us to contextualize our results in the expected conditions of the ISM. We concluded that the sole \ce{CO + OH} reaction is insufficient for the formation of \ce{CO2} on ices and that Reaction \ref{eq:abstraction} is the most promising candidate for the follow-up reaction. Considering this, is it justified to consider a small activation energy for the \ce{OH + CO -> CO2 + H} reaction in astrochemical models of molecular clouds and prestellar cores? In light of our simulations, we consider that there are at least four different cases.

\begin{enumerate}
    \item High coverage of \ce{H2O} ice and high abundance of H atoms.
    \item High coverage of \ce{H2O} ice and low abundance of H atoms.
    \item High coverage of \ce{CO} ice and high abundance of H atoms.
    \item High coverage of \ce{CO} ice and low abundance of H atoms.
\end{enumerate}

On \ce{H2O} ice (Cases 1 and 2 above), the formation of the HOCO complex is facile and does not require any energy input, with a fast reaction occurring thanks to the adsorption energy (or a fraction of it) on water ice. Moreover, the dominance of \ce{H2O} in the early stages of a molecular cloud's life, during the translucent cloud phase \citep{Snow2006}, ensure mild temperature conditions (15--50 K) that allow for diffusion of CO molecules, and relatively low extinction (A$_{v}\sim$ 1-2 mag). Under these conditions, Case 1 is the most likely one, with H atoms produced from photodissociation of \ce{H2O} and other hydrogenated molecules both in the gas and on the grain. Other mechanisms, such as cosmic ray ionization, also contribute to these fragmentation processes. Under these conditions, we determine that considering a null or low activation barrier for Reaction \ref{reac:1} in astrochemical models is justified because the H atom will ensure prompt conversion of HOCO to \ce{CO2} through reaction \ref{eq:abstraction}. However, we warn that \ce{HC(O)OH} abundance could be underestimated following this approach. At higher extinctions, but without enough CO surface coverage (Case 2, molecular cloud stage), the abundance of H atoms on grain surfaces will be reduced, and the HOCO complex will survive longer on the grain. Under these conditions, we recommend differentiating Reaction \ref{reac:1} and \ref{eq:abstraction}. 

The next two cases (Cases 3 and 4) can be treated conjointly. Our simulations show that forming the HOCO radical from \ce{CO + OH} is not straightforward on CO ice and requires initial energy input. While the energy required to initiate the reaction is not very high, the very low temperatures where Cases 3 and 4 would dominate (dense prestellar cores with T=10 K) discard the thermal energy as the initiator of the reaction. This energy input can come from a neighbouring chemical reaction because \ce{H2O} photodissociation should be a small factor in CO ices. Therefore we consider that the approach presented in \citet{Garrod2011} of modelling the \ce{CO2} formation as the three-body reaction, e.g. \ce{H + O + CO} is a good compromise to model the reaction on CO ice. Whether the three-body reaction can be coarse-grained to yield \ce{CO2 + H} directly or HOCO (and later proceed through reaction \ref{eq:abstraction}) is likely to depend on the H atom abundance. For example, an important factor should be the local cosmic ray ionization rate ($\zeta$) determining the dissociation of \ce{H2} into 2H, thus the ratio of HOCO radicals to H atoms. We must emphasize that coarse-graining the formation of \ce{CO2} through the title reaction to study \ce{CO2} formation and evolution may be acceptable only when H atom abundance overwhelms HOCO abundance. However, in doing so, the abundance of other HOCO-derived molecules like \ce{HC(O)OH} will be underestimated. Precaution is advised when the target of the models involves these molecules.

Finally we would like to discuss other possible scenarios. One possibility is that the excited formation of OH leads to  non-thermal diffusion out of the reaction site or its desorption (notice that the latter would be more plausible on CO ices due to the lower binding energy), in these cases the reaction would not take place. Another possible scenario regards the energy dissipation after HOCO is formed. Because of the high exothermicity of the \ce{CO + OH -> HOCO} reaction and the low binding energies of these radicals on CO ice, there is the possibility that HOCO chemically desorbs, or triggers the desorption of a nearby ice CO molecule. In addition, if these reactions would have to take place in the inner layers of the ice, one must take into account that energy dissipation would be even more efficient due to the larger number of intermolecular interactions and the higher number of surrounding molecules, rendering each reaction step less and less efficient.
 
%\include{tickz/disc_figure.tex}
%\begin{figure*}
%    \centering
%    \includegraphics[width=0.6\linewidth]{figures/disc_figure.png}
%    \caption{Chemical network explored in this work. Blue lines are those reactive channels active on water ice, and red-dashed ones are those on CO ice.}
%    \label{fig:network}
%\end{figure*}

\section{Conclusions} \label{sec:conclusions}

Using accurate quantum chemical calculations and microcanonical kinetic modelling, we found that the \ce{CO + OH -> CO2 + H} reaction, which has been considered as the most important producer of interstellar \ce{CO2}, is rather inefficient, and its occurrence cannot be taken for granted. The reaction proceeds through a rather stable intermediate, HOCO, and more specifically through its two structural isomers t-HOCO and c-HOCO. On \ce{H2O} ice, the formation of HOCO is feasible, but its evolution to \ce{CO2} requires a further reaction step that most likely involves H abstraction through reaction \ref{eq:abstraction}. On CO ice, we found, for the first time, that the formation of HOCO is not as efficient as currently assumed, owing to the lower adsorption energy of OH and CO molecules on CO ice. We indicate that non-thermal effects are necessary to form HOCO, and thus \ce{CO2}, on CO ice. This limitation may be behind the recent ice observations showing higher fraction of \ce{CO2} found in water-dominated environments \citep{Boogert2015, McClure2023} when comparing with apolar (CO-dominated) ices.

Because our calculations assume an ideal energy redistribution in an infinitely short time after the reactions, our results represent a lower bound for the production of \ce{HOCO} and \ce{CO2} from the \ce{CO + OH} reaction. We aim to improve the description of energy dissipation in forthcoming works to resolve ambiguous cases. We encourage further experimental work on the topic, especially on CO ices following \citet{Qasim2019}. Nonetheless, with our results, we were able to provide atomistic insight into the formation of \ce{CO2}, one of the most important interstellar ice constituents, and indicate the cases where coarse-graining of the \ce{CO + OH} reaction in astrochemical models is, to a first approximation, acceptable and not.

\begin{acknowledgements}
G.M. thanks the Japan Society for the Promotion of Science (JSPS International Fellow P22013, and Grant-in-aid 22F22013) for its support. The authors acknowledge support by the Research Center for Computational Science in Okazaki, Japan (Projects: 22-IMS-C301, 23-IMS-C128), the state of Baden-Württemberg through the bwHPC consortium and the German Research Foundation (DFG) through grant no INST 40/575-1 FUGG (JUSTUS 2 cluster) (Project: 22-IMS-C301). Y.A. acknowledges support by  Grant-in-Aid for Transformative Research Areas (A) grant Nos. 20H05847.
\end{acknowledgements}

% WARNING
%-------------------------------------------------------------------
% Please note that we have included the references to the file aa.dem in
% order to compile it, but we ask you to:
%
% - use BibTeX with the regular commands:
%   \bibliographystyle{aa} % style aa.bst
%   \bibliography{Yourfile} % your references Yourfile.bib
%
% - join the .bib files when you upload your source files
%-------------------------------------------------------------------

\bibliographystyle{aa}
\bibliography{aanda.bib}

\begin{appendix}
\section{Gas-phase comparison with \citet{Ma2012}} \label{sec:appendix1}

\begin{table}
\begin{center}
\caption{Comparison of the potential energy reaction profile (e.g. no zero-point energy corrected) between our computational method and the high accuracy studies of \citet{Ma2012}. The energy convention follows the same as in the preceding paper, considering the origin of energies in the deepest well of the PES (t-HOCO). All values are presented in kcal mol$^{-1}$ }
\label{tab:benchmark}
\begin{tabular}{lcc}
\toprule
 Structure & $\Delta$E (This Work) & $\Delta$E (\citet{Ma2012}) \\
\bottomrule
\ce{OH + CO} & 28.3 & 29.6 \\
PRC & 26.2 & 27.3 \\
PRC' & 27.4 & 28.4 \\
TS1 & 28.6 & 29.0 \\
TS1' & 32.5 & 32.7 \\
t-HOCO & 0.0 & 0.0 \\
TS2 & 9.4 & 9.3 \\
c-HOCO & 2.0 & 1.8 \\
TS4 & 33.1 & 32.0 \\
TS5 & 38.4 & 38.4 \\
\ce{HCO2} & 34.7 & 16.8 \\
TS6 & 22.1 & 20.9 \\
\ce{CO2 + H} & 8.0 & 7.0 \\
\bottomrule
\end{tabular}
\end{center}
\end{table}

We compare our energetics of the \ce{CO + OH -> CO2 + H} gas-phase reaction profile at the DLPNO-CCSD(T)/CBS//MN15-D3BJ/6-31+G(d,p) level with the high-quality CCSD(T)/AVTZ results presented in \citet{Ma2012} in \Cref{tab:benchmark}. Note that the energies presented here are not ZPVE corrected, unlike in the main manuscript. We observe excellent (between 0.0--1.3 kcal mol$^{-1}$) deviations between methods, e.g. chemical accuracy, for all structures except \ce{HCO2}. As we introduced in the methods section, this intermediate and the associated entrance and exit transition states, TS5 and TS6, are irrelevant to the reaction kinetics or dynamics \citep{Ma2012,masunov_catalytic_2018}. Hence, a wrong prediction of the energetics of this intermediate does not affect our results, and we do not include it in our kinetic simulations. Yet, it is interesting to mention the reason for the discrepancy.

In \citet{Ma2012}, the authors show that the \ce{HCO2} intermediate belongs to the C$_{2v}$ symmetry point group at the CCSD(T)/AVTZ level of theory. However, the geometries at the MN15-D3BJ/6-31+G(d,p) level converge to a C$_{s}$ intermediate. The T$_1$ diagnostic at the DLPNO-CCSD(T)/cc-pVTZ level of theory for the \ce{HCO2} intermediate hints at a strong multireference character (T$_1$=0.068), so it is not clear if the CCSD(T) or the MN15-D3BJ calculations are better in predicting the correct \ce{HCO2} geometry. However, it is clear that a dual-level approach like DLPNO-CCSD(T)/CBS//MN15-D3BJ/6-31+G(d,p) will fail due to the mismatch of geometries.  Despite the discrepancy found for \ce{HCO2}, the excellent agreement for all the relevant parts of the PES indicate that the studies on the \ce{H2O} and \ce{CO} clusters will yield the correct energetics for the system.

\end{appendix}
\end{document}